\documentclass[aps,prl,showpacs,twocolumn,groupedaddress]{revtex4}
\usepackage{graphics}
\usepackage{graphicx}
\usepackage{epsfig}
\usepackage{pstricks}
\usepackage{color}
\usepackage{amsmath}
\usepackage{amssymb}
\usepackage{ulem}
\newcommand \beq{\begin{eqnarray}}
\newcommand \eeq{\end{eqnarray}}
\newcommand \bea{\begin{eqnarray}}
\newcommand \eea{\end{eqnarray}}
\newcommand \kvec{{\bf k}}

\newcommand \Rvec{{\bf R}}

\newcommand \Gvec{{\bf G}}

\newcommand \Q{{\cal Q}}
\newcommand\rvec{{\bf r}}
\newcommand\thetavec{\vec{\theta}}
\def\simge{\mathrel{%
       \rlap{\raise 0.511ex \hbox{$>$}}{\lower 0.511ex \hbox{$\sim$}}}}
\def\simle{\mathrel{
       \rlap{\raise 0.511ex \hbox{$<$}}{\lower 0.511ex \hbox{$\sim$}}}}

\begin{document}

\title{The momentum distribution of the homogeneous electron gas}

\author{Markus Holzmann$^{1,2}$, Bernard Bernu$^2$, Carlo Pierleoni$^3$, Jeremy McMinis$^4$, David M. Ceperley$^4$, Valerio Olevano$^5$, and Luigi Delle Site$^6$}
\affiliation{$^1$Univ. Grenoble 1/CNRS, LPMMC UMR 5493, Maison des Magist\`{e}res, 38042 Grenoble, France}
\affiliation{$^2$ LPTMC, UMR 7600 of CNRS, UPMC, Jussieu, Paris, France}
\affiliation{$^3$Physics Department, University of L'Aquila,
Via Vetoio, 67100 L'Aquila, Italy}
\affiliation{$^4$Dept. of Physics and NCSA, U. of Illinois at
Urbana-Champaign, Urbana, IL 61801, USA}
\affiliation{$^5$Institut N{\'e}el, Grenoble, France}
\affiliation{$^6$Max-Planck-Institute for Polymer Research,
Ackermannweg 10, D 55021 Mainz Germany}

\date{\today}

\begin{abstract}
We calculate the off-diagonal density matrix of the homogeneous
electron gas at zero temperature using unbiased
Reptation Monte Carlo for various densities and
extrapolate the momentum distribution, and the kinetic and
potential energies  to the thermodynamic limit. Our results on
the renormalization factor allows us to validate approximate
$G_0W_0$ calculations concerning quasiparticle properties over
a broad density region ($1 \le r_s \lesssim 10$) and show that
near the Fermi surface, vertex corrections and self-consistency
aspects almost  cancel each other out.
\end{abstract}


\maketitle

The uniform electron gas (jellium) is one of the most
fundamental models for understanding electronic properties in
simple metals and semiconductors. Knowledge of its ground state
properties, and, in particular, of modifications due to
electron correlation are at the heart of all approximate
approaches to the many-electron problem in realistic models.
Quantum Monte Carlo methods (QMC) \cite{egas} have provided the
most precise estimates of the correlation energy, electron pair
density and structure factor of jellium; basic quantities for
constructing and parameterizing the exchange-correlation energy
used in density functional theory (DFT) \cite{dft}.

Correlations modify the momentum distribution, $n_k$, of
electrons, and introduce deviations from the ideal Fermi-Dirac
step-function. The magnitude of the discontinuity at the Fermi
surface ($k_F$), the renormalization factor $Z$, quantifies the
strength of a quasi-particle excitation \cite{Nozieres} and
plays a fundamental role in Fermi liquid and many-body
perturbation theory (GW) for spectral quantities. Whereas the
momentum distribution (as well as other spectral information)
is inaccessible in current Kohn-Sham DFT formulations, the
reduced single-particle density matrix -- the Fourier transform
of $n_k$ in homogeneous systems -- is the basic object in the
so-called density-matrix functional theory \cite{dmft}; these
theories rely on knowledge of $n_k$ of jellium. Inelastic x-ray
scattering measurement of the Compton profile of solid sodium
\cite{Na} have determined $n_k$, but experiments for elements
with different electronic densities are less conclusive.

In this paper, we calculate  $n_k$ for the electron gas
(jellium) by QMC in the density region $1\le r_s \le 10$. Here,
$r_s=(4 \pi na_B^3/3)^{-3}$ is the Wigner-Seitz density
parameter, $n$ is the density, and $a_B=\hbar^2/me^2$ is the
Bohr radius. In contrast to previous calculations \cite{Ortiz},
our calculations are based on more precise backflow (BF) wave
functions \cite{BF}, and a careful extrapolation to the
thermodynamic limit \cite{FSE,Momk2D}. Similar to the worm
algorithm in finite temperature path-integral and lattice Monte Carlo
\cite{Massimo,Saveriopure}, 
we have extended Reptation Monte Carlo (RMC)
\cite{Saverio} to include the off-diagonal density matrix in
order to obtain an \textit{unbiased} estimator of the momentum
distribution \cite{burkhard,Massimo2}. From our extrapolation scheme,  we derive the
exact behavior of $n_k$  close to the Fermi surface. By
comparing the renormalization factor, $Z$, with different
approximate GW theories, we can judge the importance of
self-consistency and vertex corrections within these
approaches. The excellent agreement of our QMC results with
$G_0W_0$ over a broad density region indicate strong
cancellations of vertex and self-consistency corrections  close
to the Fermi surface.

\begin{figure}
\includegraphics[width=0.48\textwidth]{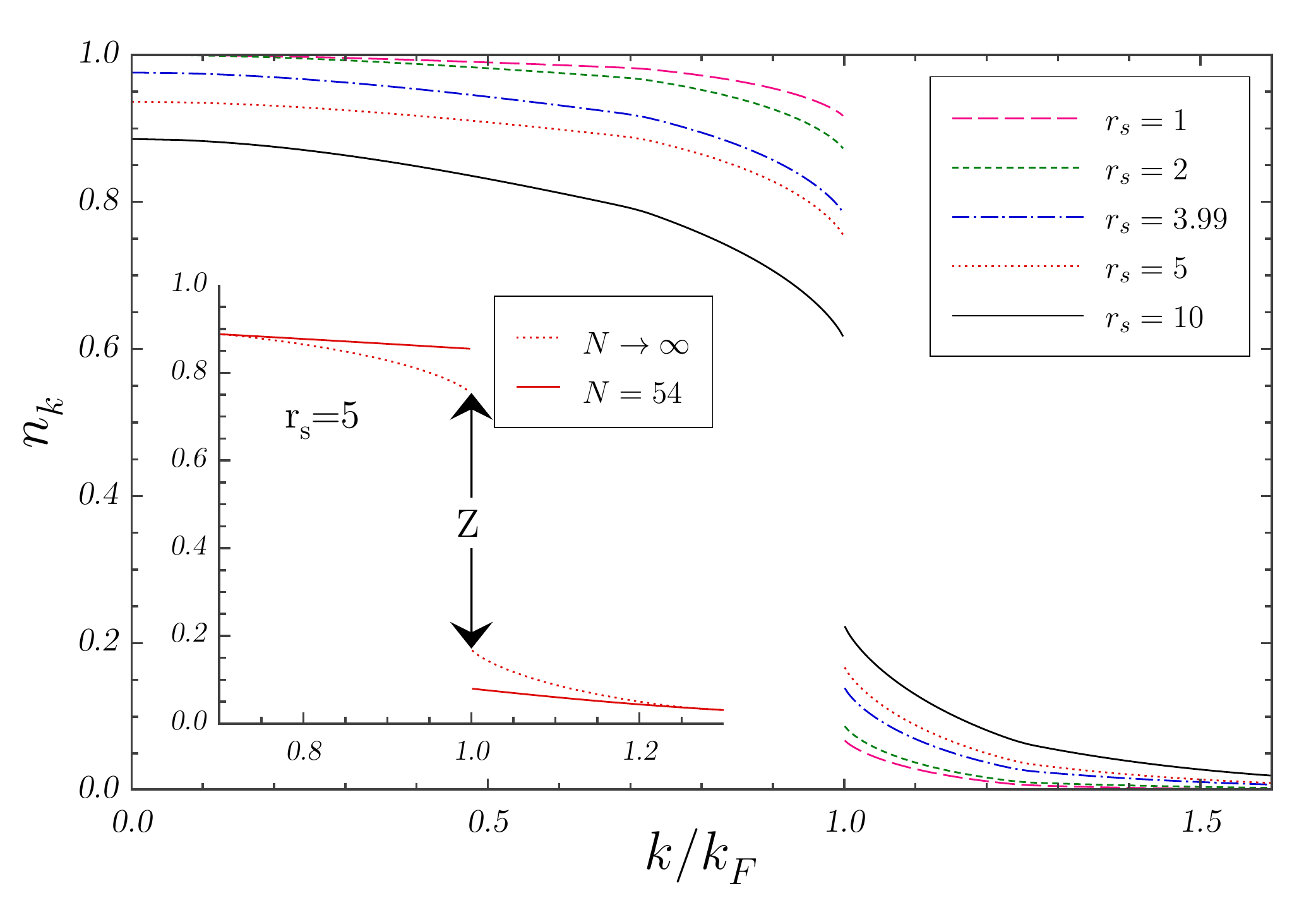}
\caption{The momentum distribution ($n_k$) of the unpolarized
electron gas for various densities extrapolated to the
thermodynamic limit. The inset shows the extrapolation of $n_k$
for $r_s=5$ from a system with $N=54$ electrons to the
thermodynamic limit, $N \to \infty$,leading to a significant
reduction of the renormalization factor $Z$.} \label{MOMK}
\end{figure}

Within Variational Monte Carlo (VMC), the ground state wave
function is approximated by a trial wave function,
$\Psi_T(\Rvec)$,  whereas within projector Monte Carlo methods,
e.g. Diffusion Monte Carlo (DMC) or reptation Monte Carlo
(RMC), the trial state is improved using $\Psi_\beta \propto
\exp[-\beta H ] \Psi_T$; this converges exponentially fast to
the true ground state for increasing projection time $\beta$.
To circumvent the so-called Fermion sign problem, calculations
are done within the fixed-node approximation, introducing small
systematic deviations from the exact Fermion ground state
\cite{Lubos}. Whenever the (approximate) nodes of the system
are described by a determinant of single particle orbitals
$\phi_n(\rvec)$, the (fixed-node) ground state wave function,
$\Psi_N(\Rvec)$, of $N$ particles at positions $\Rvec \equiv \{
\rvec_i\}$, can be written as
 \begin{eqnarray}
 \Psi_N(\Rvec)=D_N
 \exp\left[ - U_N \right]
 , \quad D_N= \det_{nl} \phi_n \left( \rvec_l + \nabla_l W_N \right)
 \label{PsiN1}
 \end{eqnarray}
where $W_N$ and $U_N$ are generalized backflow and Jastrow
potentials\cite{He3} respectively.

\begin{table}
\begin{tabular}{|c|c|c|c|c|c|c|} \hline
 $r_s$  &  1 &  2  & 3.99 & 5  & 10\\\hline
E & 1.173(2) & 0.0039(1) & -0.1555(1)& -0.1520(1) & -0.1071(1) \\
T & 2.290(3) & 0.6024(5) & 0.1688(1) &  0.1131(1) & 0.0349(1) \\
V &  -1.116(1) & -0.5985(1) & -0.3243(1)&  -0.2651(1)& -0.1421(1) \\
g(0)& 0.268(3) &   0.152(2)& 0.057(2)  & 0.034(1) & 0.0036(4) \\
\hline
$n_0$ & 0.999& 0.998 & 0.97& 0.93 & 0.88\\
$n_2$ & 0.038& 0.066&  0.12 &0.098 & 0.21\\
$\bar{n}$ & 0.490 & 0.477& 0.460& 0.456& 0.414 \\
 \hline
\end{tabular}
\caption{The total ($E$), potential ($V$) and kinetic energy
($T$) per particle in $Ry$, and the contact value of the pair
correlation function $g(0)$, all extrapolated to the
thermodynamic limit from unbiased RMC calculations with
backflow (BF) nodes. We further give parameters of the momentum
distribution at small $k$ ($n_0$, and $n_2$) $n(k \to
0)=n_0-n_2 (k/k_F)^2$, and at $k_F$:
$\bar{n}=[n(k_F^+)+n(k_F^-)]/2$. } \label{tablezero}
\end{table}
\begin{table}
\begin{tabular}{|c|c|c|c|c|c|c|} \hline
 $r_s$  &  1 &  2  & 3.99 & 5  & 10\\\hline
  BF-RMC &  0.84(2) & 0.77(1) & 0.64(1) & 0.58(1) & 0.40(1)
  \\
\hline
 SJ-VMC &  0.894(9) & 0.82(1) & 0.69(1) & 0.61(2) & 0.45(1)
 \\
  BF-VMC &  0.86(1) & 0.78(1) & 0.65(1) & 0.59(1) &  0.41(1)
  \\
  \hline
  \hline
  $G_0W_0$ \cite{Hedin} & 0.859  & 0.768 & $0.646^*$   & 0.602  &  0.45\\
$GW_0$ \cite{GW0} & & 0.804 &  $0.702^*$  & & \\
$GW$ \cite{GW} & & 0.846 &$ 0.793^*$  & &\\
Lam \cite{Lahm}& 0.896 & 0.814 &$ 0.615^*$  & 0.472 & \\
RPA\cite{Lahm} & 0.843 &  0.700 & $0.442^*$  & 0.323 &\\
SJ-DMC \cite{Ortiz} & 0.952 & 0.889 & & 0.725 &0.593 \\
\hline
\end{tabular}
\caption{Renormalization factor, $Z$, extrapolated to the
thermodynamic limit from unbiased  RMC calculations with
backflow  nodes (BF-RMC), together with SJ-VMC, and BF-VMC
results, compared with perturbative results from literature
(literature values$^*$ are at $r_s=4$ instead of $r_s=3.99$).
Previous SJ-DMC results \cite{Ortiz} used \textit{mixed}
estimators without thermodynamic limit extrapolation. }
\label{tableone}
\end{table}

From an approximate ground state wavefunction, $\Psi_N(\Rvec)$,
we obtain the reduced single particle density matrix
\cite{McMillan} \beq f_N(\rvec)=
 \left\langle F(\Rvec;\rvec) \right\rangle_N
,\quad F=\frac{1}{N} \sum_i  \frac{
\Psi_N(\Rvec:\rvec_i+\rvec)}{\Psi_N(\Rvec)} \label{fr} \eeq
where $\Rvec:\rvec_i+\rvec$ indicates that  the position of
particle $i$ is displaced by $\rvec$, and $\langle \dots
\rangle_{N} \equiv  \int d \Rvec \dots |\Psi_N|^2/Q$ with
$Q\equiv \int d \Rvec  |\Psi_N|^2$ playing the role of a
partition function. The Fourier transform of $f_N(\rvec)$
directly yields the momentum distribution, $n_\kvec^N$, of the
electrons  per spin \beq n_\kvec^N=\frac{1}{2V}\int d \rvec
e^{-i\kvec\cdot \rvec} f_N(\rvec) \eeq where $V$ is the volume.

The large variance of the estimator of the off-diagonal density
matrix, Eq.~(\ref{fr}), makes precise calculations very
time-consuming. To reduce the variance for homogeneous systems
with plane wave orbitals: $\phi_n(\rvec) \propto e^{i \kvec_n
\cdot \rvec}$, we separate the ideal gas density matrix,
$f_{id}(\rvec)=\sum_n \phi_n^*(\rvec)\phi_n(0)/ \sum_n
|\phi_n(0)|^2$, based on the estimator \beq
F_{id}(\Rvec;\rvec)=\frac{1}{N} \sum_i  \frac{
 D_N(\Rvec:\rvec_i+\rvec;W_N(\Rvec))}{D_N(\Rvec;W_N(\Rvec))}
 \label{Fid}
\eeq where the determinants  on the r.h.s. of
Eq.~(\ref{Fid})  are evaluated using the backflow coordinates,
$W_N(\Rvec)$, of the diagonal configuration $\Rvec$ with
un-displaced particle coordinates. Expanding it around
$\rvec=0$, we can explicitly verify that $f_{id}(\rvec) =
\left\langle F_{id}(\Rvec;\rvec) \right\rangle_N$, so that the
$F-F_{id}$ is a reduced variance estimator\cite{Bloch} of the
difference: $f_N-f_{id}$.

There is a problem with projecting methods to calculate
properties other than the energy. Forward walking or
reweighting methods based on using $\Psi_\beta$ in
Eq.~(\ref{fr}), become very inefficient for long projection
time, since the variance increases exponentially with $\beta$.
To avoid this problem, mixed estimators, based on $\Psi_\beta
\Psi_0$, are frequently used but they can introduce a
systematic bias. Unbiased estimators for the pair correlation
function, potential and kinetic energy have been obtained
within RMC \cite{Saverio}. Based on a generalized partition
function, $\Q$, we extend RMC to include sampling of
off-diagonal matrix elements \cite{Saveriopure}
\begin{align} \Q &= \int \! d \Rvec \,  |
\Psi_{\beta/2}(\Rvec) |^2& \nonumber
\\
&+\! \!
\frac{s}{N} \sum_i  \!\!   \int \frac{ d\rvec}{V}
\! \! \int_0^\beta  \frac{d \tau}{\beta}\! \!   \int \! \! d \Rvec
 |\Psi_{\beta \! -\! \tau} (\Rvec) \Psi_\tau (\Rvec:\rvec_i+\rvec)|
\end{align}
where  $s$ is a parameter  used to optimize the efficiency
($s=0$ corresponds to the usual diagonal RMC \cite{Saverio}).
Similar to the worm-algorithm used in continuous Path-integral
calculations \cite{Massimo}, our calculations include moves
which ``open'' (or ``close'') a path from diagonal space
$\Rvec$ to off-diagonal space $(\Rvec,\rvec_i+\rvec)$. Such
moves are included at $\tau=0$ and ``propagated'' by reptation
moves \cite{Saverio,bounce} to the interior of the path
($\tau>0$). In contrast to previous calculations using
so-called mixed estimators \cite{Ortiz}, this generalization
gives an \textit{unbiased} estimator of the off-diagonal
density matrix, $f_N(\rvec)$, and the momentum distribution,
$n_k^N$. Reduction of the variance based on the considerations
above, Eq.~(\ref{Fid}), is still possible, but less effective.

Quantum Monte Carlo results are obtained for typically $N
\lesssim 10^3$ electrons. The extrapolation to the
thermodynamic limit introduces important quantitative and
qualitative changes of the momentum distribution around the
Fermi surface, $k_F$  \cite{Momk2D}. For a homogeneous periodic
system, the orbitals are plane waves: $\phi_n(\rvec)=\exp[i
(\kvec_n+\thetavec)\cdot \rvec]$, in the Slater determinant of
Eq.~(\ref{PsiN1}), where $\kvec_j \in \Gvec_N \equiv  \{
(n_1,n_2,n_3) 2 \pi V^{-1/3} \}$ with integer $n_i$, and
$\thetavec$ can be chosen to introduce twisted boundary
conditions \cite{TBC,FSE}. For a normal Fermi liquid, we
further have $|\kvec_j+\thetavec|\le k_F$, and the generalized
backflow and Jastrow potential $W_N$ and $U_N$ can be written
exclusively in terms of collective coordinates $\rho_\kvec
=\sum_n e^{i \kvec \cdot \rvec_n}$ and their gradients
\cite{BF,He3}. Using the wavefunction ``potentials'', $W_N$ and
$U_N$, expressed as continuous functions in terms of the
collective coordinates, the relation between the wave function
in the limit $N \to \infty$ to a finite system is well defined,
as it just amounts to evaluations on a denser grid in
$\kvec$-space \cite{FSE,Momk2D}.

Let us first discuss the finite size scaling for a Slater-Jastrow (SJ)
wave function: a determinant with $W_N\equiv 0$, together with
a two-body Jastrow correlation, $U_N=\sum_k u_k \rho_k
\rho_{-k}/2V$. We further assume that the function $u_k$ is
analytically given.
In our SJ-VMC calculations, we use the Gaskell form
$2nu_k^{SJ} \equiv -S_0^{-1}(k)+\left[S_0^{-2}(k)+ 2n v_k/
\varepsilon_k \right]^{1/2}$ where $S_0(k)$ is the ideal gas
structure factor, $v_k=4 \pi e^2/k^2$, and
$\varepsilon_k=\hbar^2 k^2/2m$ \cite{Gaskell,SJanalytic}.  Neglecting
mode-coupling between single particle modes in $D_N$ and
collective modes described by $U_N$, the single particle
density matrix, Eq.~(\ref{fr}), can be approximated as
\begin{equation}
f_{N}(\rvec)
 \approx
   f_c(\rvec) \equiv
  \left\langle \frac{ D_{N}'
 }{D_{N}}
  \right\rangle_N
  \left\langle
  e^{-(U'_{N}-U_{N})} \right\rangle_N
\end{equation}
where the prime indicates the off-diagonal configuration, e.g.
$D_{N}' \equiv D_{N}(\Rvec:\rvec_1+\rvec)$. Within the cumulant
and rotating wave approximation, we then obtain an explicit
expression,
\begin{eqnarray}
f_c(\rvec)
 &\simeq&  f_{id}(r)  \exp\left[- x_N(r) \right]
\label{cumulant}
\\
x_N(r)&=&
\frac{1}{V}  \! \sum_{|\kvec|\le k_c}
 \left[
 u_k \left( S_k \!- \!1 \right)
 +n u_k^2 S_k \right]
 \left[e^{i \kvec \cdot \rvec} \!-\! 1 \right]
\label{x}
\end{eqnarray}
where $S_k=\langle \rho_{\kvec} \rho_{-\kvec}\rangle_N/N$ is
the structure factor, $f_{id}(r)=2 \sum_{k\le k_F} e^{i \kvec
\cdot \rvec}/N$ is the single particle density matrix of the
corresponding ideal gas, and we have neglected contributions of
short wave length modes, $k_c \approx 0.48 r_s^{1/2} k_F$
\cite{BP}. Further, we can use $S_k \approx [2n u_k +
1/S_0(k)|^{-1}$ to express $S_k$ in terms of $u_k$ and
$S_0(k)$, which is based on  assuming  gaussian statistics for
$\rho_k$, so than Eq.~(\ref{cumulant}) gives an explicit
expression for $f_N (\rvec)\approx f_c(\rvec)$ in terms of a
given Jastrow factor. Whereas the resulting model,
Eq.~(\ref{cumulant}), depends weakly on $k_c$, so that $f_N(r)$
and $n_k$ are only qualitatively described, the
size-extrapolation is quantitatively correct, as it is dominated
by the Jastrow singularity $u_k \to
(v_k/2n\varepsilon_k)^{1/2}$ and $S_k \to
(2nv_k/\varepsilon_k)^{-1/2}$  for $k\to 0$ stemming from the
plasmon contributions.

Since we expect that mode-coupling is negligible in the long
wave length limit, the cumulant expression,
Eq.~(\ref{cumulant}), can be used to determine the size
corrections of QMC calculations of the finite system
\begin{equation}
f_\infty(\rvec)= \frac{2}{n} \int \frac{d^3k }{(2\pi)^3}
n_k^N e^{i \kvec \cdot \rvec}
e^{-(x_\infty(r)-x_N(r))}. \label{finfty}
\end{equation}
Here $n_k^N$ is the momentum distribution of the $N$
electron system,
defined for all values of $\kvec$ in a grand canonical ensemble
using twisted boundary conditions
\cite{FSE}.
From the Fourier transform of $f_\infty(\rvec)$,
Eq.~(\ref{finfty}), we obtain the extrapolated momentum
distribution, $n_k^\infty$. A related linearized expression has
been used to extrapolate the momentum distribution of the
two-dimensional electron gas in Ref.~\cite{Momk2D}.

Following the analysis of Ref.~\cite{Momk2D}, leading order
corrections to the renormalization factor,
$Z_N=n_{k_F-}^N-n_{k_F+}^N$, are given by
\begin{eqnarray}
Z_\infty& \simeq &Z_N \exp[-\Delta_N]
\\
 \Delta_N &=& \int_{-\pi/L}^{\pi/L} \frac{d^3q}{(2\pi)^3} \frac{u_q}{2}
 \left[ 1+ {\cal O} \left( [2n u_q S_0(q) ]^{-1} \right) \right]
 \nonumber
\\
&=& c \left( \frac{3}{4\pi} \right)^{1/3} \left( \frac{r_s}{3} \right)^{1/2} N^{-1/3}
+ {\cal O} \left( N^{-2/3} \right)
\nonumber
\end{eqnarray}
where $c\simeq 1.221$ is a numerical factor to account for the
cubic integration volume \cite{MomkRPA}. Whereas the asymptotic region is only
reached for large systems with $N^{1/3} r_s^{1/2}  \gg 1$, the
extrapolation based on the full expression, Eq.~(\ref{finfty}),
includes corrections beyond the leading order term. Analyzing
Eq.~(\ref{finfty}) around $k_F$,  we obtain the \textit{exact}
leading order behavior with an infinite slope at $k_F$
\begin{align}
&n(k \to k_F^{\pm})  \simeq
n(k_F^\pm)
\nonumber \\
& \quad+ \frac{Z_\infty}{2\pi} \left(\frac{9\pi}{4} \right)^{1/3} \sqrt{\frac{r_s}{3}} \left[ \frac{k}{k_F}-1 \right]
\log \left| \frac{k}{k_F}-1 \right|. \label{nkf}
\end{align}

Size extrapolation, discussed above, requires the knowledge of
the structure factor, $S_k$, and the Jastrow potential, $u_k$,
in Eq.~(\ref{x}). The QMC calculation of the $N$-particle
system allows us only to determine them  on a finite grid in
$\kvec$ space, but the analytic continuation to the dense grid
can be done by interpolation from their known behavior at small
k \cite{FSE}. Whereas $S_k$ can be calculated directly,
$u_k=u_k^{SJ}$ is only known explicitly for VMC calculations
using a Slater-Jastrow trial function. In general, imaginary
time projection and backflow introduce an effective Jastrow
potential, $u_k$, different from the explicitly given form of
the underlying trial wavefunction. Expecting small changes at
long wave length, $u_k=u_k^{SJ}+\delta u_k$, we  obtain the
modifications $\delta u_k$ from from changes in the structure
factor $\delta S_k=S_k-S_k^{SJ}$ by linear response. For our
purpose, mode coupling can be neglected, as well as deviations
from gaussian statistics, so that $\delta S_k/\delta u_{k'}
\simeq - 2n S_k^{-2} \delta_{\kvec,\kvec'}$ for $k \to 0$.
Therefore, the effective Jastrow factor of wave functions
including backflow and projection can be determined from the
structure factor.

Using  SJ-VMC calculations with $u_k^{SJ}$ for $N=54$ to
$N=1024$ electrons, we have checked that size extrapolations
based on Eq.~(\ref{finfty}) with $N=54$ are reliable. Thus, the
more expensive backflow VMC and RMC calculations based on the
analytical wave functions in Ref.~\cite{BF} are only done with
that size.  Extrapolated results on the total energy $E$,
unbiased estimators from reptation for the potential ($V$) and
kinetic energies ($T$), and the contact value of the pair
correlation function, $g(0)$, are given in
table~\ref{tablezero}. The momentum distribution is shown in
Fig.~\ref{MOMK}. The values for the renormalization factor,
$Z$, together with different perturbative results from the
literature are given in table~\ref{tableone}.
Table~\ref{tablezero} also contains the values of the momentum
distribution at the origin, $n_0$, the negative slope at the
origin, $n_2$, and $\bar{n}=(n_{k_F^-}+n_{k_F^+})/2$. These
values can be used to parameterize the momentum distribution
along the lines given in Ref.~\cite{paramet}, together with
$Z$, the exact large $k$ asymptotics \cite{cusp}, $n(k\to
\infty) = (9/2) r_s^2 g(0)/k^8$,  and the exact behavior close
to the Fermi surface, Eq.~(\ref{nkf}). Whereas the mixed
estimator usually employed in DMC calculations, introduces a
small bias in the momentum distribution, size extrapolation
introduces large systematic modifications which limit the
precision of the calculations. Previous DMC results
\cite{Ortiz},  using mixed estimators and SJ nodes, suffer from
these strong finite size effects and  overestimate $Z$ by a
large amount.

In summary, we have calcuated the momentum distribution using a
new unbiased and much more accurate Monte Carlo method, and
extrapolated the results to the thermodynamic limit. In
particular, our data allows a quantitative comparison of the
renormalization factor, $Z$, with approximate calculations (see
table~\ref{tableone}). The excellent agreement of our results
with $G_0W_0$ \cite{Rice,Hedin, LocalField} over the whole
metallic density region $r_s \lesssim 5$, strongly indicates
that vertex corrections and self-consistency issues -- neither
is included in $G_0W_0$ -- are canceling each other, at least
close to the Fermi surface.

Computer time at CNRS-IDRIS is acknowledged,
Projet IDRIS 061801. CP is supported by IIT under the SEED project grant n 259 SIMBEDD.


\begin{thebibliography}{99}
\bibitem{Bloch} This variance reduction
   can be generalized to periodic systems with Bloch orbitals, $\phi_{n\kvec}(\rvec) \propto e^{i\kvec \cdot \rvec} \varphi_n(\rvec)$, displacing only
   the phase in the determinant, $\phi_{n\kvec}(\rvec_i) \to e^{i\kvec \rvec} \phi_{n\kvec}(\rvec_i)$ when $\rvec_i \to \rvec_i+\rvec$.
\end{thebibliography}
\end{document}